# Fatigue-resistant high-performance elastocaloric materials via additive manufacturing

Huilong Hou [1], Emrah Simsek [2], Tao Ma [2], Nathan S. Johnson[3], Suxin Qian [4], Cheikh Cissé [3],

Drew Stasak [1], Naila Al Hasan [1], Lin Zhou [2], Yunho Hwang [5], Reinhard Radermacher [5], Valery

I. Levitas [2, 6], Matthew J. Kramer [2, 7], Mohsen Asle Zaeem [3], Aaron P. Stebner [3], Ryan T. Ott [2],

Jun Cui [2, 7], Ichiro Takeuchi [1, *]

[1] Department of Materials Science and Engineering, University of Maryland, College Park, Maryland 20742, United States of America

[2] Division of Materials Science and Engineering, Ames Laboratory, Ames, Iowa 50011, United States of America

[3] Department of Mechanical Engineering, Colorado School of Mines, Golden, Colorado 80401, United States of America

[4] Department of Refrigeration and Cryogenic Engineering, Xi'an Jiaotong University, Xi'an, Shaanxi 710049, People's Republic of China

[5] Center for Environmental Energy Engineering, Department of Mechanical Engineering, University of Maryland, College Park, Maryland 20742, United States of America

[6] Departments of Aerospace Engineering and Mechanical Engineering, Iowa State University, Ames, Iowa 50011, United States of America

[7] Department of Materials Science and Engineering, Iowa State University, Ames, Iowa 50011, United States of America

* Correspondence and requests for materials should be addressed to I.T. (takeuchi@umd.edu).

(Manuscript was updated on August 12$^{nd}$, 2019)

# Abstract

Elastocaloric cooling, which exploits the latent heat released and absorbed as stress-induced phase transformations are reversibly cycled in shape memory alloys, has recently emerged as a frontrunner in non-vapor-compression cooling technologies. The intrinsically high thermodynamic efficiency of elastocaloric materials is limited only by work hysteresis. Here, we report on creating high-performance low-hysteresis elastocaloric cooling materials via additive manufacturing of Titanium–Nickel (Ti–Ni) alloys. Contrary to established knowledge of the physical metallurgy of Ti–Ni alloys, intermetallic phases are found to be beneficial to elastocaloric performances when they are combined with the binary Ti–Ni compound in nanocomposite configurations. The resulting microstructure gives rise to quasi-linear stress-strain behaviors with extremely small hysteresis, leading to enhancement in the materials efficiency by a factor of five. Furthermore, despite being composed of more than 50% intermetallic phases, the reversible, repeatable elastocaloric performance of this material is shown to be stable over one million cycles. This result opens the door for direct implementation of additive manufacturing to elastocaloric cooling systems where versatile design strategy enables both topology optimization of heat exchangers as well as unique microstructural control of metallic refrigerants.

**One Sentence Summary**: 3D printing produces highly efficient solid-state cooling nanocomposites with long fatigue life.

# Introduction

The first-order transitions of caloric (magnetocaloric, mechanocaloric, and electrocaloric) materials(*1-3*) can be exploited for giant cooling effects, but hysteresis is their Achilles heel since it represents work lost in every heat-pumping transformation cycle resulting in dissipated heat, and it can ultimately lead to materials fatigue and failure. In fact, while long-life fatigue properties are critical for applications of caloric materials, they are only occasionally reported on.

Elastocaloric cooling, one of the mechanocaloric cooling mechanisms, makes use of the reversible martensitic transformations of shape memory alloys (SMAs) to induce an adiabatic change in temperature, $\Delta T$, (or isothermal change in entropy, $\Delta S$) by absorption and release of transformation enthalpy (*4*). With $\Delta T$ as large as 17 K(*5*) and $\Delta S$ up to 70 J kg$^{-1}$ K$^{-1}$ (*6*), the energy saving potential of elastocaloric cooling technology has been widely recognized by the community working on non-vapor compression cooling technologies (*7*). Functioning elastocaloric cooling prototypes with over 100 W in cooling capacity(*8*) as well as elastocaloric regenerative heat pumps with temperature span larger than 19 K(*9, 10*) have been demonstrated. However, thermomechanical hysteresis that limits the efficiency of their thermodynamic performances as well as their fatigue behaviors remains a concern.

In this work, we report on successful laser directed-energy-deposition (L-DED) synthesis of elastocaloric alloys composed of intermetallic and alloy phases arranged in nanocomposite microstructures. We take advantage of L-DED, where metal powders are melted locally and solidified rapidly(*11, 12*), to synthesize nanocomposites consisting of transforming, elastocaloric binary Ti-Ni alloy and a non-transforming TiNi$_3$ intermetallic phase in a two-phase mixture of comparable volume fractions, with intricate dendritic structures. This unique configuration enlists the non-transforming intermetallic phase for biasing the phase transformation leading to considerable improvement in elastocaloric efficiency as well as reversibility of the transformation through minimizing the work hysteresis.

Owing to the stress transferring mechanism built into the nanocomposite microstructures, the L-DED alloys exhibit substantially reduced hysteresis with a quasi-linear stress-strain behavior resulting in a remarkable five-fold increase in the materials efficiency defined as the ratio of materials coefficient of performance ($COP_{\text{materials}}$) to Carnot $COP$. We show that the elastocaloric thermodynamic cycle of these materials is stable over more than a million cycles. In contrast to rate-dependent hysteresis commonly observed in traditionally processed SMAs (*13, 14*), the hysteresis of the L-DED SMAs is nearly rate-independent (from 0.0002 s$^{-1}$ to 0.2 s$^{-1}$), facilitating high-frequency elastocaloric operations. We use a constitutive model and *in situ* synchrotron diffraction experiments to confirm that their unique properties originate from kinematics of load transfer between transforming and non-transforming phases.

# Results

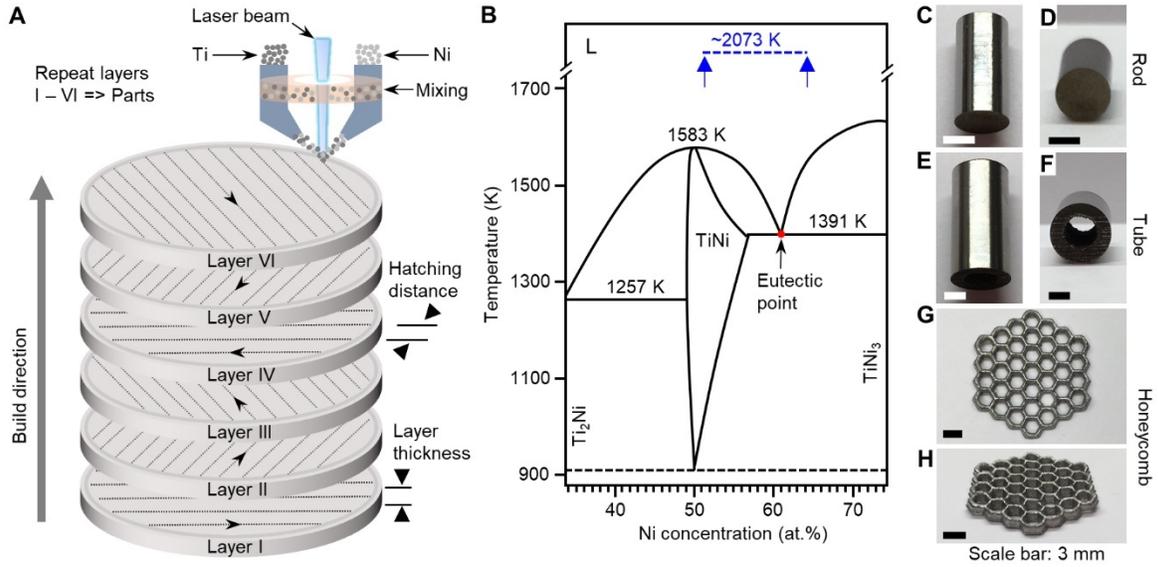

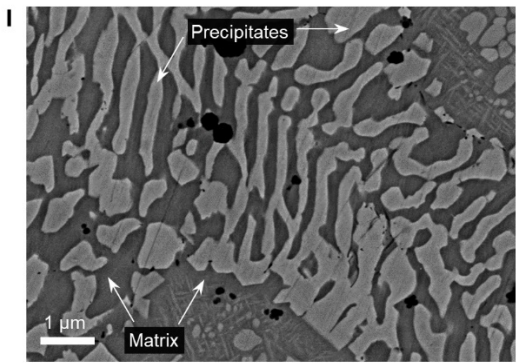
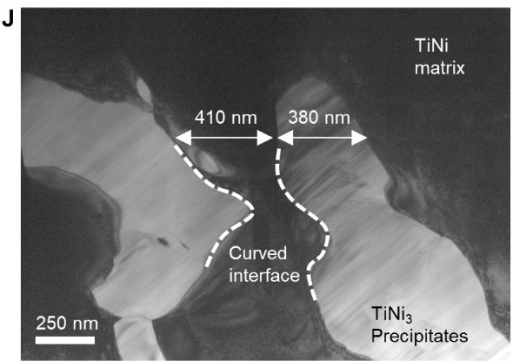
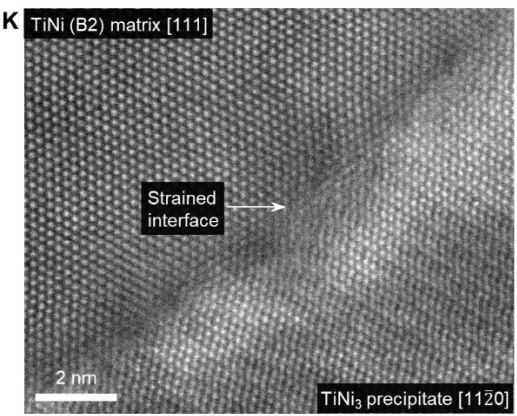
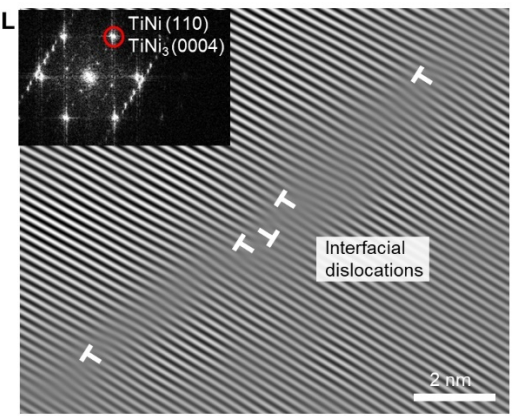

**Fig. 1. Design of elastocaloric nanocomposite alloys by directed energy deposition.** (**A**) Schematic representation of a laser directed-energy-deposition (L-DED) process. Flows of Ni and Ti powders are individually controlled. The Ni and Ti powders are mixed and then fed to the laser beam. An induced molten pool moves to build materials layer by layer with prescribed parameters (Fig. S1). (**B**) Phase diagram of binary Ti–Ni alloys (adapted with permission from *Binary alloy phase diagrams* © 1990 ASM International) highlighting in blue the Ni-rich composition near a eutectic point and the molten pool temperature ~2,073 K used in this work. Rapid cooling of the localized molten pool leads to nanocomposite alloys. (**C**)–(**H**) Photographs of L-DED produced nanocomposite Ti–Ni rods, tubes, and honeycombs in top (C),(E),(G) and front (D),(F),(H) views, respectively. (**I**)–(**K**) SEM image (I), bright-field TEM image (J), and high-resolution HAADF-STEM image (K) of as-built $Ti_{48.5}Ni_{51.5}$ nanocomposite alloy. In (I), the regions with different contrasts are crystallographically identified to be TiNi and $TiNi_3$ phases (Fig. S2 and S3). In (J), typical curved interfaces between TiNi and $TiNi_3$ phases which provide the unique stress-transfer are delineated. In (K) (a zoomed-in view of a curved interface), TiNi and $TiNi_3$ phases have an orientation relationship of $TiNi_{[111]} \parallel TiNi_{3[11\bar{2}0]}$, although each is slightly off the zone axis due to lattice strains within the interface (Fig. S4). (**L**) Inverse fast Fourier transform (IFFT) image from the circled spot in the FFT image (inset; generated from (K)). Interfacial dislocations are identified and marked with T symbols. The dislocations play the role of pre-existing sites of high nucleation potency for martensitic forward and reverse transformations.

The key features of the laser directed-energy-deposition (L-DED) process (shown in Fig. 1A) are a millimeter-scale molten pool of mixed powders and a rapid cooling rate of more than $10^3$ K s$^{-1}$(*15*). Metal nanocomposites made by, for example, casting(*16*) can display a stress-transfer mechanism responsible for high strength, a desirable attribute of functional alloys. Since eutectic solidification can naturally lead to the formation of composites, the eutectic point in the Ni-rich composition range (Fig. 1B) of binary Ti–Ni was used to obtain elastocaloric nanocomposite alloys using L-DED(*17*). Optimization of processing parameters (such as layer thickness, hatching space) was guided by a normalized processing map(*18*) for high denseness (≈99%) and mechanical integrity, and the molten pool temperature in operation was maintained to be 1,973–2,173 K, as measured *in situ* by a ThermaViz pyrometer. Different compositions of Ti–Ni alloys were printed by adjusting the ratio of the flow rate of elemental Ni and Ti powders. Fig. 1C–H display some of the printed geometries.

Rapid cooling of the molten pool during L-DED enables precipitation from off-eutectic compositions in a volume fraction comparable to that of eutectic structures, as predicted by Scheil model(*15, 19*). Here, we observe a substantial amount of precipitates in a wide compositional range of the Ti–Ni alloys produced by L-DED (Fig. 1B). Curved microstructures can nucleate and grow, because the temperature gradient (highest at center and lowest at periphery) of the molten pool leads to circulation of mass and heat within the pool driven by Marangoni shear stress(*20*), thereby creating local perturbations of solute concentration and equilibrium temperature(*21*) on solid–liquid interfaces and breaking up the plane front in growth of steady-state eutectics. As a result of non-equilibrium conditions, a typical microstructure of L-DED produced Ti–Ni alloys consists of transforming TiNi and non-transforming TiNi$_3$ phases with large aspect ratios, curved interfaces, and comparable volume fractions (Fig. 1I). The size scale of the microstructure is inversely proportional to the cooling rate($19$), which is at least two orders of magnitude higher in L-DED than that of casting (~0.1 K s$^{-1}$) leading to a mixture of two phases at a submicrometer scale (Fig. 1J).

Large curvatures of the interfaces between the cubic B2-ordered TiNi phase and the hexagonal D0$_{24}$-ordered TiNi$_3$ phase (Fig. 1J) in the nanocomposite microstructures can be naturally accommodated with small lattice mismatches to make their interfaces semi-coherent. An atomic-scale view of the adjacent regions displays strained boundaries (Fig. 1K) where interfacial dislocations are located (Fig. 1L). Pre-existing sites of high nucleation potency such as dislocations have been reported to trigger atomic shearing for nucleation of martensite($22$) where a nucleation energy barrier is lowered (or completely suppressed in the case of spontaneous growth($23$)). These interfacial dislocations inherent to the curvatures and additional dislocations induced by mechanical pre-treatment (Fig. S5) therefore serve as pre-existing nucleation sites to reduce energy barriers for martensite during the forward transformation and for austenite during the reverse transformation. In addition, these same nucleation sites can act as "micro-pockets" to accommodate remnant austenite and martensite after forward and reverse transformations, respectively, thereby eliminating the necessity of barrier-overcoming stage for nucleation during cyclic loading. After proper self-organization, pre-straining, and pre-stressing (shakedown state, Fig. S5), the intricate nanoscale network of connected microstructure suppresses the dislocation motion($24$) and limits transformation dissipation resulting in enhanced cyclic stability.

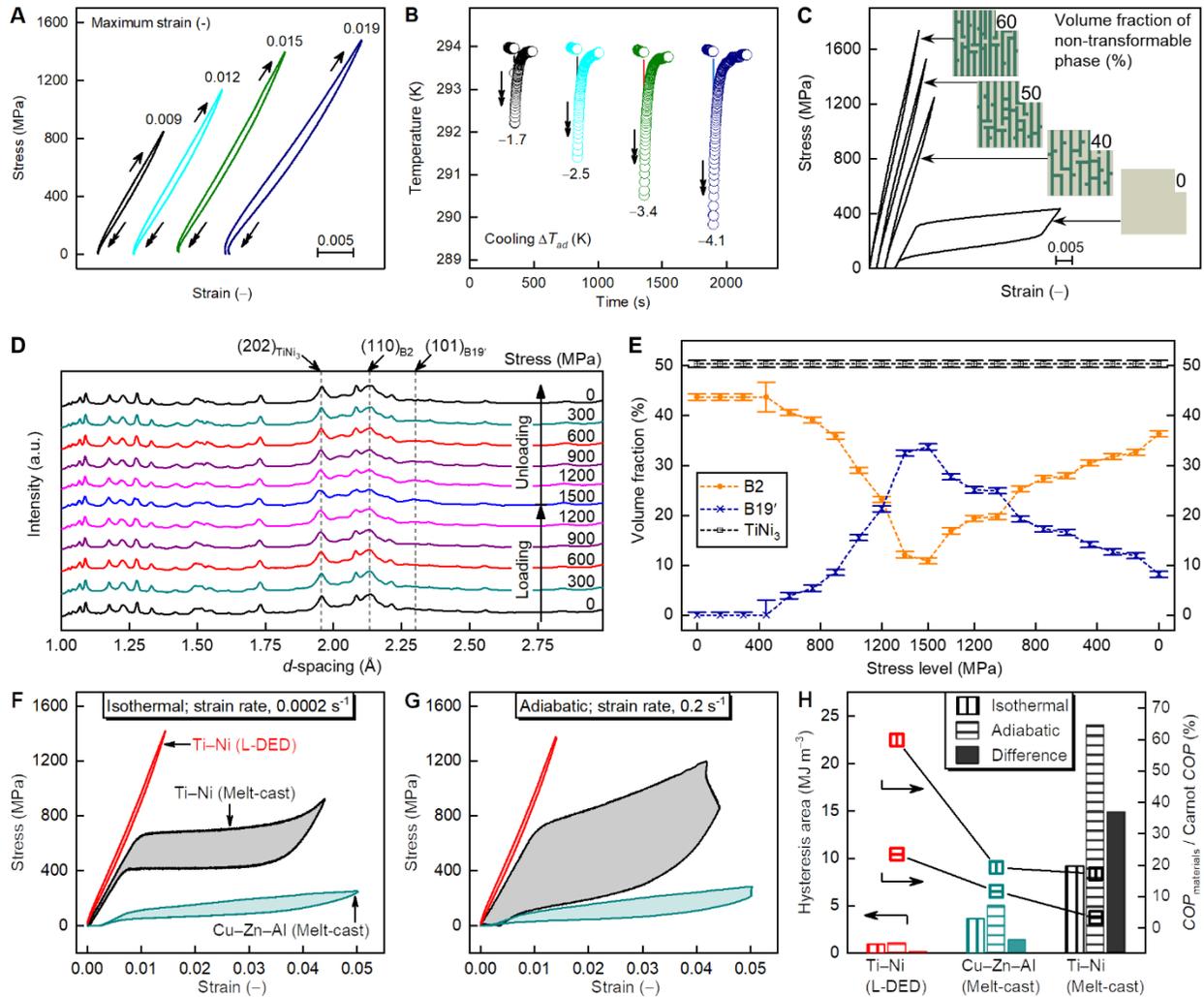

**Fig. 2. Unusual properties of elastocaloric nanocomposite alloys fabricated by laser directed-energy-deposition.** (**A** and **B**) Measured stress-strain curves (A) and the corresponding elastocaloric cooling at room temperature (B) of L-DED produced $Ti_{48.5}Ni_{51.5}$ nanocomposite alloys aged at 923 K for 3 hours. The single arrows in (A) denote loading and the double arrows in (A) and (B) correspond to unloading. (**C**) Simulated stress-strain curves from a micromechanics model that accounts for the volume fraction of non-transforming phase (insets). (**D** and **E**) Synchrotron X-ray diffraction patterns during *in situ* loading-unloading (D) and the determined volume fraction of primary phases at different stress levels during the cycle (E). (**F** and **G**) Comparison of stress-strain curves for $Ti_{48.5}Ni_{51.5}$ nanocomposite alloys and melt-cast $Ti_{49.2}Ni_{50.8}$ and $Cu_{68}Zn_{16}Al_{16}$ alloys at the strain rate of 0.0002 s$^{-1}$ for isothermal loading/unloading (F) and 0.2 s$^{-1}$ for adiabatic loading/unloading (G). In (F) and (G), the area enclosed by the loading/unloading curves represents total dissipation energy per unit volume associated with hysteresis. (**H**) Comparison of hysteresis area under isothermal and adiabatic loading/unloading as well as the ratio of $COP_{materials}$ to Carnot $COP$ for L-DED nanocomposite alloys and melt-cast alloys. The color code for each material is common for (F)–(H).

The L-DED nanocomposite alloys exhibit quasi-linear behaviors and substantially reduced hysteresis (Fig. 2A). The full strain recovery upon unloading is accompanied by a cooling $\Delta T_{\text{ad}}$ (Fig. 2B), a signature of martensitic transformation, which reaches 4.1 K.

Herein, the quasi-linear recovery behavior arises from the load transfer between the non-transforming, stiff intermetallic phase and the transforming non-load-bearing phase. The effective modulus of the L-DED nanocomposite alloys (~80–90 GPa) is higher than the typical austenite (~50–60 GPa): the non-transforming intermetallic $TiNi_3$ phase is stiffening the alloy. As a result, as the austenite transforms to martensite, the intermetallic phase continues to carry the load elastically, and the resulting overall behavior is quasi-linear. To confirm this mechanism, we have simulated the crossover from a regular superelastic to quasi-linear behavior by varying the volume fraction of non-transforming intermetallic phase and observed the appearance of quasi-linear behavior at a level of 40%, 50%, and 60% (Fig. 2C).

The small hysteresis observed here is due to the topology- and defect-controlled kinematics of numerous nucleation events and coalescence, where spatially dispersed pre-existing nucleation sites (Fig. 1L) favor continual, heterogeneous nucleation of new martensite followed by their coalescence. The resulting volumetric densities of obstacles that austenite-martensite transformation fronts meet in the course of transformation are reduced and require a decreased amount of frictional work to overcome, as observed in Cu–Zn–Al alloys(*25*). Additionally, the intermetallic phase has a large volume fraction (~50%), and it effectively guides the transformation process through elastic interaction with the transforming phase. This process, in turn, tempers multiple instabilities occurring during traditional nucleation and fast growth and reduces energy dissipation and effective interfacial friction. The progression is captured in *in situ* synchrotron diffraction measurements (Fig. 2D, 2E).

The commonly-observed rate-dependent hysteresis (e.g., the difference in hysteresis curves between Fig. 2F and 2G) is attributed to transformation-induced heat in SMAs where surface convection dominates heat transfer. From an explicit integral equation of the specific dissipated energy $\Delta E$ (which is equal to the generated heat) (*26*), we can approximate $\Delta E$ as:

$$\Delta E \cong E_{\text{fr}} + \Delta T_{\text{ad}} \cdot \Delta s \tag{1}$$

where $E_{\text{fr}}$ is the irreversible specific energy which is the generated heat through interface friction, $\Delta T_{\text{ad}}$ is the adiabatic change in temperature, and $\Delta s$ is the specific entropy change associated with the phase transformation. The $\Delta E$ during a stress-strain cycle manifests itself as the hysteresis area (divided by density), and it increases with enlarged hysteresis. This relation can also explain the nearly rate-*independent* hysteresis observed here in nanocomposite alloys (Fig. 2H) where thermal conduction (thermal conductivity $\approx$ 18 W m$^{-1}$ K$^{-1}$) through a large volume fraction of non-transforming phase and surface convection (with convective heat transfer coefficient $\approx$ 4 W m$^{-2}$ K$^{-1}$) collectively facilitate effective heat transfer and rejection in a transformation cycle. In this instance, the second term on the right of Eq. (1) becomes considerably small due to the rate of heat dissipation approaching the rate of heat generation.

Decreasing $E_{\text{fr}}$ contributes to additional reduction in $\Delta E$. In fact, $E_{\text{fr}}$ consists of two components: $E_{\text{fr}} = E_{\text{f}} + E_{\text{p}}$ (*27*), where $E_{\text{f}}$ is the heat dissipated from frictional work in a transformation cycle and $E_{\text{p}}$ is the heat dissipated by plastic work within austenite-martensite interfaces due to their coherency loss. Although friction is ubiquitous in the propagation of austenite-martensite interfaces(*28*), reducing extended interfacial motions by having uniformly distributed sites for nucleation and coalescence can substantially curtail frictions, leading to reduced $E_{\text{f}}$. The resultant minimization of $E_{\text{f}}$ accounts for the substantial reduction in $E_{\text{fr}}$ (Fig.

2H). In other alloy systems, relaxing local strain energy associated with phase transformation via improving lattice compatibility was found to lead to significant reduction in $E_\text{p}$ (*29-31*).

Thermodynamics of cooling devices dictates that isothermal loading/unloading in Stirling-like cycles can naturally lead to high efficiencies due to their inherently small hysteresis (*8, 32*). However, Stirling-like operation cycles require much longer time per cycle (leading to reduced output wattage) and additional system components for effective heat transfer (*32*). In comparison, adiabatic loading/unloading in Brayton-like cycles(*33*) can operate much faster with relatively simple heat-exchange systems, albeit suffering from lower intrinsic efficiency due to the larger hysteresis (Fig. 2G). $COP_\text{materials}$ in Brayton-like cycles are governed by the directly measured $\Delta T_\text{ad}$ with the adiabatic hysteresis, and $COP_\text{materials}$ in Stirling-like cycles are regulated by the latent heat with the isothermal hysteresis, based on a thermodynamically derived equation with full work recovery (See Methods). In both cycles, the hysteresis of L-DED nanocomposite alloys is extremely small and has a negligible difference (indicating rate-independency). With a Carnot $COP = 37.5$ for $T_\text{h} = 308$ K and $T_\text{c} = 300$ K, the ratio of $COP_\text{materials}$ to Carnot $COP$ of L-DED nanocomposite alloys is approximately 5 times that of melt-cast counterparts (Fig. 2H).

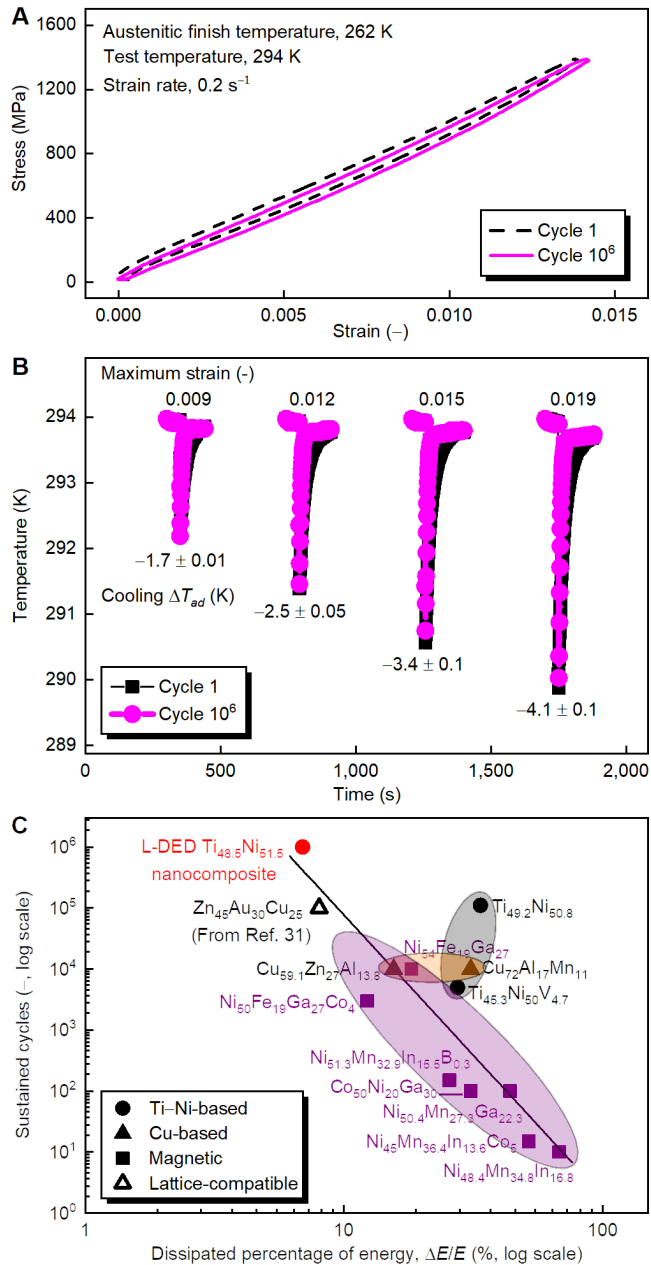

**Fig. 3. Stability of Ti–Ni nanocomposite alloys over one million compression cycles and comparison to other reported bulk elastocaloric materials.** (**A** and **B**) Compressive stress-strain curves (A) and elastocaloric cooling (B) of laser directed-energy-deposition (L-DED) produced $Ti_{48.5}Ni_{51.5}$ nanocomposite alloys aged at 923 K for 3 hours before and after one million cycles. The nanocomposite alloys can sustain one million cycles with no change in mechanical and elastocaloric properties. (**C**) Log–log plot of the dissipated fraction of input energy, $\Delta E/E$, versus sustained compressive cycles for bulk elastocaloric materials in this work as well as those reported in the literature. A dissipated fraction of energy is the ratio of hysteresis area, $\Delta E$, in a transformation cycle to the input energy, $E$. "Lattice-compatible" refers to the alloy where the lattice parameters of transformed and untransformed phases exhibit exceptional lattice compatibility (*31*). The straight line is a linear fit. The data from both polycrystalline and

single-crystal materials are included. The numerical values used in this plot as well as strain amplitude and references are listed in Table S1.

We have studied the long-term stability of the L-DED nanocomposite alloys. As seen in Fig. 3A and 3B, we find that the nanocomposites are stable in their mechanical behavior and elastocaloric response for over 1 million cycles, indicating that they can be used in regular commercial products with a typical ten-year life (operating at < 1 Hz(*8*)). Small hysteresis is one important factor responsible for the observed long-term stability of alloys: previously, we have shown that by tuning the lattice compatibility using stoichiometry in ternary alloys, one can minimize hysteresis of martensitic transformation and improve its reversibility to extended numbers of cycles (*29, 34*). However, comparisons of different SMA materials reveal that the absolute value of hysteresis is not the only determining factor. In fact, magnetic SMAs such as polycrystalline Ni–Mn–In and Ni–Fe–Ga (Table S1) seem to deteriorate quickly after a small number of cycles (~100) even with a hysteresis area as small as 1.2 MJ m$^{-3}$. It is known that for stress-induced fatigue, the endurance limit (that is, the stress amplitude able to attain a prescribed number of cycles, usually $10^7$, at zero mean stress) is proportional to the ultimate strength of materials by a factor of ≈0.33 (*35*). As can be seen in Fig. 3C, across a spectrum of elastocaloric materials, it is the *ratio* of hysteresis area $\Delta E$, to the input work, $E$, which ultimately determines the number of cycles that the materials can sustain their performance over.

To understand this trend, we consider an analogy to the well-known *S–N* concept conceived by Wöhler in 1858(*36*) that connects the stress amplitude (*S*) to the cycles to failure (*N*) in structural fatigue of materials and obtain a correlation of $\Delta E/E$ (hysteresis as a fraction of input energy) to the cycles to "functional failure", *N*, (which we define as the number of cycles at the onset of loss of their functionality) in the log–log plot (Fig. 3C). In an ideal case of $\Delta E/E = 0$ (i.e., transformation with no hysteresis), the number of cycles to functional failure would asymptotically approach infinity. SMAs typically exhibit hysteresis in superelastic cycles; the best compounds hitherto reported for cycling are $Zn_{45}Au_{30}Cu_{25}$ alloys optimized through tuning the lattice parameters(*31*) and $Ti_{48.5}Ni_{51.5}$ nanocomposite alloys with friction-limited kinematics in this work, both of which possess an $\Delta E/E$ less than 10%. Because of similarity in the hysteresis behavior associated with input work among different materials, the energy-based ($\Delta E/E$)–*N* correlation observed here for elastocaloric materials could, in principle, apply to other caloric materials (i.e., magnetocaloric and electrocaloric materials). Even though the data on fatigue behavior of other caloric materials are somewhat limited (Table S1), our preliminary analysis indicates that the same correlation holds for them as well. Caloric materials based on first-order transitions with reported low cyclability (e.g., <10,000 cycles) can potentially have their functional fatigue lives extended if their $\Delta E/E$ can be decreased by, for instance, materials processing.

The conventional wisdom in the SMA community is that presence of non-equiatomic Ti–Ni phases such as $TiNi_3$ in the TiNi matrix is detrimental to materials integrity as the presence of brittle phases precipitated along grain boundaries can lead to fracture from local stress concentration(*37*) and mismatch stress generated by transformation-induced shape distortions in neighboring grains (*38*). The non-equiatomic phases have also plagued the self-propagating high-temperature synthesis used for porous Ti–Ni for decades as they occur inevitably and produce chemical inhomogeneity in porous implants (*39*). Here, for the first time, we have created a Ti–Ni-based elastocaloric material whose exceptional stability and unusual operational efficiency are in fact *derived from* their unique and intricate nanocomposite structures made possible by additive manufacturing.

**Acknowledgements**

Advanced Research Projects Agency-Energy (ARPA-E) of the U.S. Department of Energy (DOE) supported the original characterization of shape memory alloys at the University of Maryland under grant number ARPA-E DEAR0000131. Use of the LENS equipment was supported by the Critical Materials Institute, an Energy Innovation Hub funded by the Advanced Manufacturing Office of the Office of Energy Efficiency and Renewable Energy of the U.S. DOE. The work at Ames Laboratory was also supported by the Division of Materials Science and Engineering of the Basic Energy Sciences Programs of the Office of Science of the U.S. DOE under contract number DE-AC02-07CH11358 with Iowa State University. A.S. and C.C. acknowledge the funding from the National Science Foundation (Career Award CMMI-1454668) and N.J. acknowledges a Los Alamos National Laboratory Additive Manufacturing Graduate Fellowship. The *in situ* diffraction experiments were performed at Sector 1-ID-E of the Advanced Photon Source at Argonne National Laboratory, which is supported by the Basic Energy Sciences Programs of the Office of Science of the U.S. DOE under contract number DE-AC02-06CH11357. We thank Norman M. Wereley and Thomas Pillsbury for assistance in cyclic experiments and Peter Y. Zavalij for help with X-ray diffraction analysis.

**Author contributions**

I.T. initiated and supervised the research. H.H., J.C., and I.T. planned the experiments and designed the samples. E.S. and R.O. prepared the materials using the LENS. D.S. and N.H. characterized the composition of materials. H.H. carried out the experiments including heat treatment, DSC measurement, superelastic tests, elastocaloric cooling measurements, and long-cycle tests, and analyzed the data. T.M., L.Z., and M.K. conducted SEM and TEM analysis. N.J., C.C., and A.S. performed and analyzed the *in situ* diffraction experiments, and C.C., M.A.Z., and A.S. performed the finite element modeling. S.Q., Y.H., and R.R. discussed the thermodynamic cycles. V.L. discussed the mechanism involving interfacial dislocations, stabilization, and shakedown. H.H., A.S., J.C., and I.T. wrote the paper with substantial input from other authors. All authors contributed to the discussion of the results.

## Additional information

Supplementary Materials are available in the submission.

The authors declare no competing interests.

## Data availability

All data are available in the manuscript and supplementary materials.

**List of contents for Supplementary Materials**

Materials and Methods

Supporting online text

Figs. S1 to S8

Table S1

## Materials and Methods

Materials fabrication

Additive manufacturing of Ti–Ni alloys was carried out by using an L-DED system, Laser Engineered Net Shaping (LENS$^{TM}$) (MR-7, Optomec Inc.) equipped with a 1 kW (1,064 nm wavelength) IPG Yb-fiber laser, four-nozzle coaxial powder feeders, and a motion control system. Two powder feeders were used to separately deliver elemental Ni and Ti powders (size ~45–88 μm for Ni (purchased from American Elements) and ~45–106 μm for Ti (purchased from AP&C Advanced Powders & Coatings Inc.); purity >99.9%; gas-atomized) and the rotational speed of each feeder was used to control the mass flow rate of powders in order to tailor the mixing ratio and thus alloy composition. A laser beam with a spot size of 0.5–1.0 mm and a Gaussian intensity distribution created a molten pool on a Titanium plate substrate for flowing powders in a high-purity argon environment (< 1.0 μL L$^{-1}$ oxygen). A three-dimensional computer-aided design model was used to guide the laser paths of contour and hatch for consecutive tracks on one layer and progressive movement along the Z-direction to generate subsequent layers. Continuous scan strategy was applied with a unidirectional scanning direction. The inverse of dimensionless hatch spacing, which is beam radius divided by hatch spacing, was optimized to be 2.0–3.0 and the dimensionless volumetric energy density (required to melt the powders in a single scan) was tuned to be 1.7–4.3. The varied parameters yielded a sample density of ≈98.9%. Within a 300 mm$^3$ work envelope, cylindric parts were built with dimensionless layer thickness ~6.8 (Fig. S1) for laboratory tests, while tubular and honeycomb-shaped parts were built with dimensionless layer thickness ~0.7 as exemplified geometries.

The alloy compositions were characterized using wavelength dispersive spectroscopy (Electron Probe Microanalyzer 8900R, JEOL Inc.) with calibrated standards, after sequential polishing with a final 0.05 $\mu$m surface finish. Differential scanning calorimetry (Q100, TA

Instruments) was performed at a scanning rate of 10 K min$^{-1}$ per F2004−05 ASTM standard. Post-fabrication heat treatments(*40*) were conducted in a high-temperature tube furnace (Lindberg/Blue M, Thermo Fisher Scientific Inc.) at a heating rate of 10 K min$^{-1}$ under argon environment (Fig. S6). Melt-cast alloys of $Ti_{49.2}$–$Ni_{50.8}$ at.% were purchased from Confluent Medical Technologies Inc. and $Cu_{68}$–$Zn_{16}$–$Al_{16}$ at.% was synthesized at Ames Laboratory.

Mechanical and elastocaloric cooling testing

Uniaxial compressions were conducted on the machined specimens (10 mm in length and 5 mm in diameter) at room temperature using a servohydraulic load frame (810, MTS Systems Corp.) equipped with a load cell of 250 kN. A factory-calibrated extensometer with a gauge length of 5 mm (632.29F-30, MTS Systems Corp.) was used to record the strains. The temperature of the specimens was measured using T-type thermocouples (nominal size of 0.5 mm × 0.8 mm) attached to the middle of the specimens, recorded using a data recorder (cDAQ-9171, National Instruments Corp.), and stored using a LabVIEW program. Mechanical pre-treatment was conducted to initiate fully recoverable behaviors (Fig. S5).

Mechanical cycling tests were performed in a displacement-controlled mode with a sinusoidal loading profile at room temperature. After conversion, the nominal mean strain, $\varepsilon_m$, was set to 2.0% with a strain amplitude, $\Delta\varepsilon/2$, of 1.8% to keep the specimen subjected to compressive stress throughout the cycles. The cycle frequency was 0.05−0.1 Hz which was about the same as that of operative cycles in cooling system prototypes (*8, 41*). 1,000,000 cycles were conducted and then the materials were tested to compare with the initial state.

Microstructure characterization

A focused ion beam microscope (Helios NanoLab G3 UC, Thermo Fisher Scientific Inc.) equipped with a micromanipulator was used to prepare transmission electron microscopy (TEM) specimens by lifting out lamellae along the build direction of the materials and thinning down to ~100 nm thickness under 30 kV, followed by a sequential cleaning under 5 kV and 2 kV. Scanning electron microscopy (SEM) images were collected at an accelerating voltage of 10 kV and a working distance of 4 mm. TEM observations were performed using a probe-corrected scanning transmission electron microscope (STEM) (Titan Themis 300, FEI Company) operated under an accelerating voltage of 200 kV. High-angle annular dark-field (HAADF) STEM images were acquired in a detection range of 99–200 mrad at a probe convergence angle of 18 mrad, and the dispersive X-ray spectroscopy (EDS) spectra and maps were collected using a Super-X EDS detector.

*In situ* compression testing during X-ray diffraction

*In situ* compression testing was performed during synchrotron X-ray diffraction measurements using the third generation Rotational and Axial Motion System (RAMS3)(*42, 43*) load frame at the Sector 1-ID-E hutch of the Advanced Photon Source (APS) at Argonne National Laboratory. A 1.2 mm wide by 1 mm tall monochromatic X-ray beam with 71.6 keV energy was used to illuminate the gage of the $1\times1\times2$ mm$^3$ parallelepiped compression specimen. During both loading and unloading, at load increments of 150 MPa between 0 and 1,500 MPa compressive loads, diffraction patterns were recorded every 0.5° of sample rotation on a GE-41RT area detector(*44*) located 1,449.3 mm away from the specimen as the specimen was rotated from 0° to 360° about the loading axis.

To analyze phase fraction evolutions with loads, all images collected for each load step were summed and integrated into a single histogram, and Rietveld refinement was then performed using GSAS-II (*45*). In performing the refinements, the structures of the majority TiNi$_3$ and B2 phases were firstly used in the refinement model, allowing lattice strains and microstrains to refine for both phases. Despite averaging the diffraction data over all sample rotations about the loading axis, the data still showed signatures of texture, especially for the TiNi$_3$ phase. This texture is indicative of directional solidification and growth in L-DED processes (*15*). Then, sixth and tenth order spherical harmonics functions were used in modeling the B2 and TiNi$_3$ phases, respectively. After the majority phases were fit, the non-transforming, minority Ni and Ti$_4$Ni$_2$O phases (Fig. S7) were then added to the model. While the lattice strain and microstrain parameters were stable for the Ti$_4$Ni$_2$O phase, the microstrain for the Ni phase had to be manually adjusted and fixed. The same refinement strategy was then used for the first four loading steps (150, 300, 450, 600 MPa). The same phase fractions were determined for 0, 150, and 300 MPa loads within a fitting standard deviation. At 450 MPa, the refinement changed, indicating that B2 was transforming to B19′. To fit the martensite phase, the phase fractions of the non-transforming phases were fixed, and the B2 and B19′ phase fractions were refined against each other, in addition to lattice and microstrains for all phases, starting with the peak load (1,500 MPa), and working toward 450 MPa, for both loading and unloading data. The Rietveld model fit to the data for 0 and 1,500 MPa load, including the difference between the measured data and the Rietveld model, is visualized in Fig. S8.

Constitutive modeling

Abaqus finite element models of 1×1 mm$^2$ size with sectional thicknesses of 0.1 mm were made to mimic the aspect ratios of TiNi versus TiNi$_3$ morphologies experimentally observed in Fig. 1I. The models were meshed using approximately 21,000 4-node doubly curved S4 elements with 0.01 mm size. Elements were assigned to belong to either a transforming TiNi phase or a non-transforming phase, with phase assignments mimicking the observed microstructures as reasonable as possible considering the mesh size. The non-transforming phase was assumed to be a volume-averaged mixture of TiNi$_3$, Ti$_4$Ni$_2$O and Ni (volume fractions) according to the quantitative analysis of synchrotron X-ray diffraction patterns. More specifically, an equivalent non-transforming phase was defined with the effective Young's modulus, $\tilde{E} = 0.85 \times E_{\text{TiNi}_3} + 0.1 \times E_{\text{Ti}_4\text{Ni}_2\text{O}} + 0.05 \times E_{\text{Ni}}$ and Poisson's ratio $\tilde{v} = 0.85 \times v_{\text{TiNi}_3} + 0.1 \times v_{\text{Ti}_4\text{Ni}_2\text{O}} + 0.05 \times v_{\text{Ni}}$, where the Young's modulus and the Poisson's ratio for TiNi$_3$, Ti$_4$Ni$_2$O, and Ni are 235 GPa, 44 GPa, and 200 GPa, and 0.28, 0.35, and 0.31, respectively. Models made using 40%, 50%, and 60% volume fractions of this non-transforming phases were used in the simulations. The transforming TiNi phase was simulated using the superelastic model that is built into Abaqus with $E_{\text{TiNi}-\text{B2}} = 46$ GPa, $E_{\text{TiNi}-19'} = 28$ GPa, $v_{\text{TiNi}-\text{B2}} = 0.33$, $v_{\text{TiNi}-19'} = 0.33$, $\sigma_M^s$ (start stress for forward transformation into martensite) = 300 MPa, $\sigma_M^f$ (finish stress for forward transformation into martensite) = 500 MPa, $\sigma_A^s$ (start stress for reverse transformation into austenite) = 250 MPa, $\sigma_A^f$ (finish stress for reverse transformation into austenite) = 50 MPa, and $\varepsilon_L$ (transformation strain) = 5%.

Thermodynamic analysis

Elastocaloric materials coefficient of performance $COP_{\text{materials}}$ were computed based on the thermodynamic analysis of our custom single-stage elastocaloric testing system (*8*), where the elastocaloric materials exhibit a uniform temperature profile at $T_h$ (the temperature at hot heat exchanger) and $T_c$ (the temperature at cold heat exchanger). The elastocaloric Brayton-like cycle consists of isentropic (adiabatic) loading and unloading processes, and two heat transfer processes under constant stress fields. The elastocaloric Stirling-like cycle consists of isothermal loading and unloading processes, and two heat transfer processes under constant stress fields. By merging thermodynamics-based equations(*46*) with hysteresis-contained Equation (1), we make a universal form of $COP_{\text{materials}}$ in Equation (S1):

$$COP_{\text{materials}} = \frac{T_c \cdot \Delta s - \Delta E/2}{(T_h - T_c) \cdot \Delta s + \Delta E} \tag{S1}$$

Here, $\Delta s$ is computed using $\Delta s = q/T_c$, where $q$ is the absorbed heat, which can be obtained using $\Delta T_{ad}$ as $q = C_p \times \Delta T_{ad}$ with a specific heat capacity $C_p$ of 550 J kg$^{-1}$ K$^{-1}$ (Ti–Ni) and 420 J kg$^{-1}$ K$^{-1}$ (Cu–Zn–Al), or by $\Delta H_{M \to A}$ via $q = \Delta H_{M \to A}$. Materials densities $\rho$ are 6,500 kg m$^{-3}$ for Ti–Ni and 7,700 kg m$^{-3}$ for Cu–Zn–Al. $T_h$ and $T_c$ are set to be 308 K and 300 K, respectively, to be consistent with AHRI Standard 210/240. Here, Carnot $COP = \frac{T_c}{(T_h - T_c)} = 37.5$.

**Supporting online text. Optimization of processing parameters for alloy design.**

To optimize process parameters, we have selected a recommended processing window in a normalized processing diagram(*18, 47*). The dimensionless volumetric energy density, $E^*$, is defined in Equation (S2)(*18, 48*):

$$E^* = \frac{p^*}{v^* \cdot l^*} = \frac{A \cdot p}{2 \cdot v \cdot l \cdot r_b \cdot \rho \cdot C_p \cdot (T_m - T_0)} \tag{S2}$$

where $p^* = \frac{A \cdot p}{r_b \cdot k \cdot (T_m - T_0)}$ is the dimensionless laser power, $v^* = \frac{v \cdot r_b}{D}$ is the dimensionless laser scanning speed, $l^* = \frac{2 \cdot l}{r_b}$ is the dimensionless layer thickness, $A$ is the surface absorptivity ($\approx 0.26$)(*49*), $p$ is the laser power, $v$ is the laser scanning speed, $l$ is the layer thickness, $r_b$ is the beam radius, $\rho$ is the density, $C_p$ is the specific heat capacity, $T_m$ is the melting temperature, and $T_0$ is the initial temperature of the material. Besides, $h^* = \frac{h}{r_b}$ is the dimensionless hatch spacing. In the combinations of processing parameters, we keep $1/h^*$ to be 2.0–3.0 and $E^*$ to be 1.7–4.3.

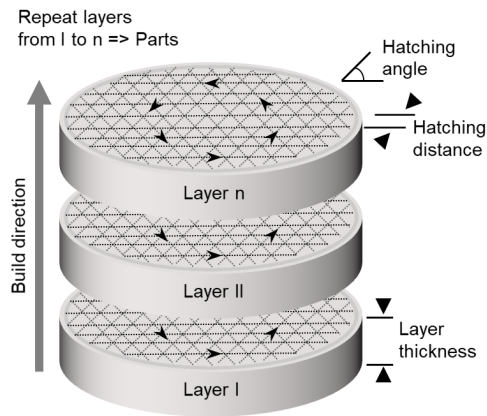

**Fig. S1. Design strategy of *in situ* thermal processing**. Schematic representation of building an extended thickness in a layer and multiple hatching on the same layer for nanocomposite alloy rods. The dimensionless layer thickness is 6.8, and the inverse of dimensionless hatch spacing is 3.0. The laser beam passes six times on each layer as the hatching angle is changed by 60° with each run. This process results in imparting intense thermal energy, similar to the multiple melting-remelting processes in the conventional melt-casting method (*50*).

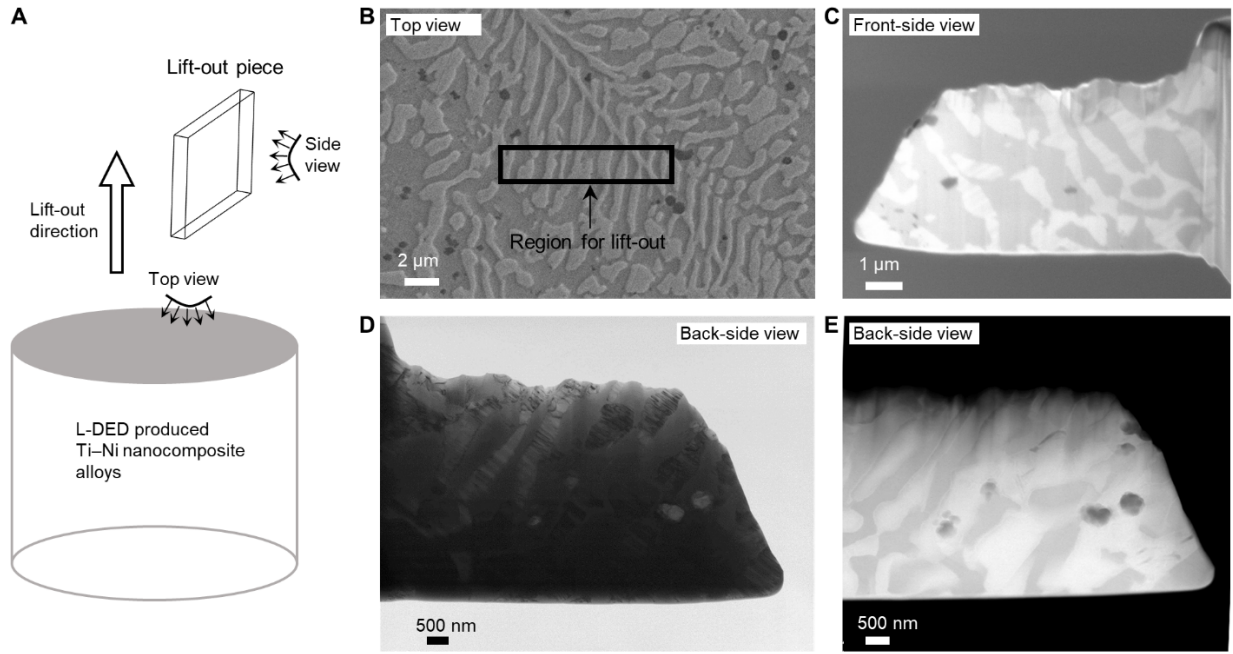

**Fig. S2. Typical microstructure morphology.** (**A**) Schematic showing the lift-out of a slice from the L-DED produced Ti–Ni nanocomposite alloys. (**B**) SEM image of the specimen surface from a top view showing the region for lifting. (**C**) SEM image of the lateral cross-section surface from a front-side view showing a co-existence of the TiNi and TiNi$_3$ phases. (**D**) TEM diffraction-contrast image from a back-side view showing the morphology and inner structure of the TiNi and TiNi$_3$ phases. (**E**) STEM image from a back-side view showing the distinct compositions of the TiNi phase versus TiNi$_3$ phase.

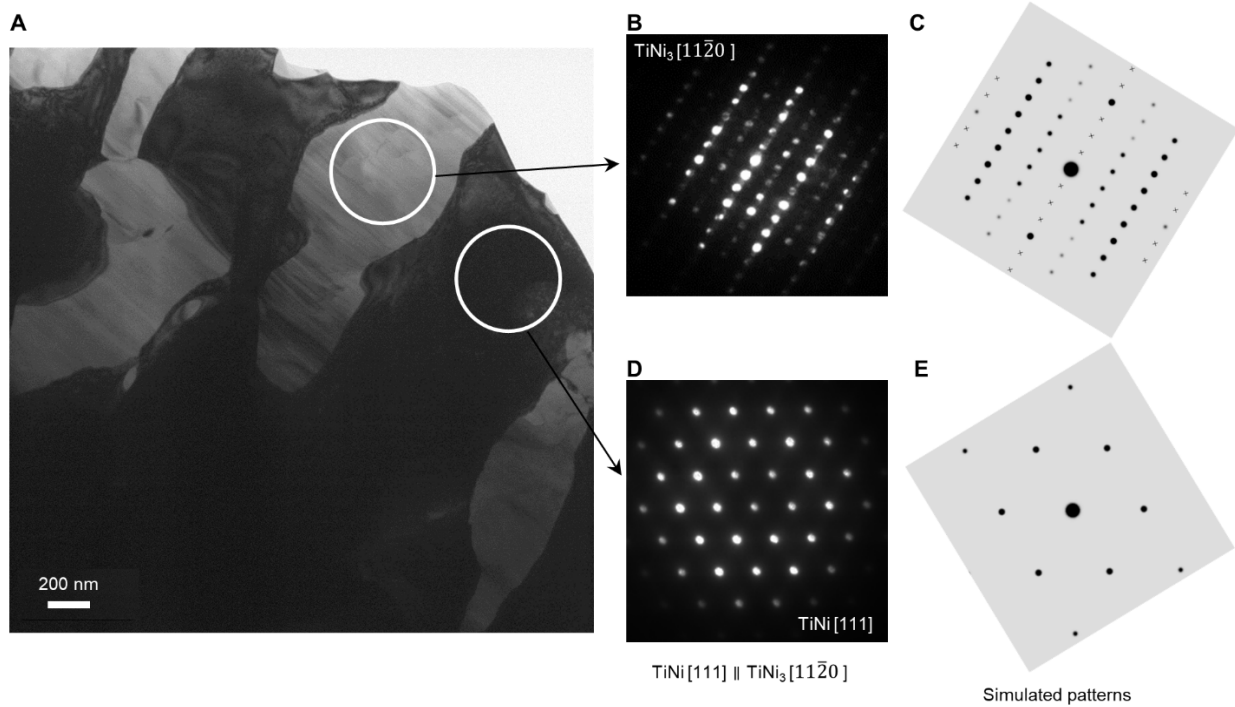

**Fig. S3. Structures analyzed by electron diffractions.** (**A**) TEM image showing the co-existence of the TiNi phase and TiNi$_3$ phase in the L-DED produced Ti–Ni nanocomposite alloys. (**B** and **C**) Selected area diffraction patterns (B) and simulated patterns (C) confirming the TiNi$_3$ phase at a zone axis of [11$\bar{2}$0]. (**D** and **E**) Selected area diffraction patterns (D) and simulated patterns (E) confirming the TiNi phase at a zone axis of [111].

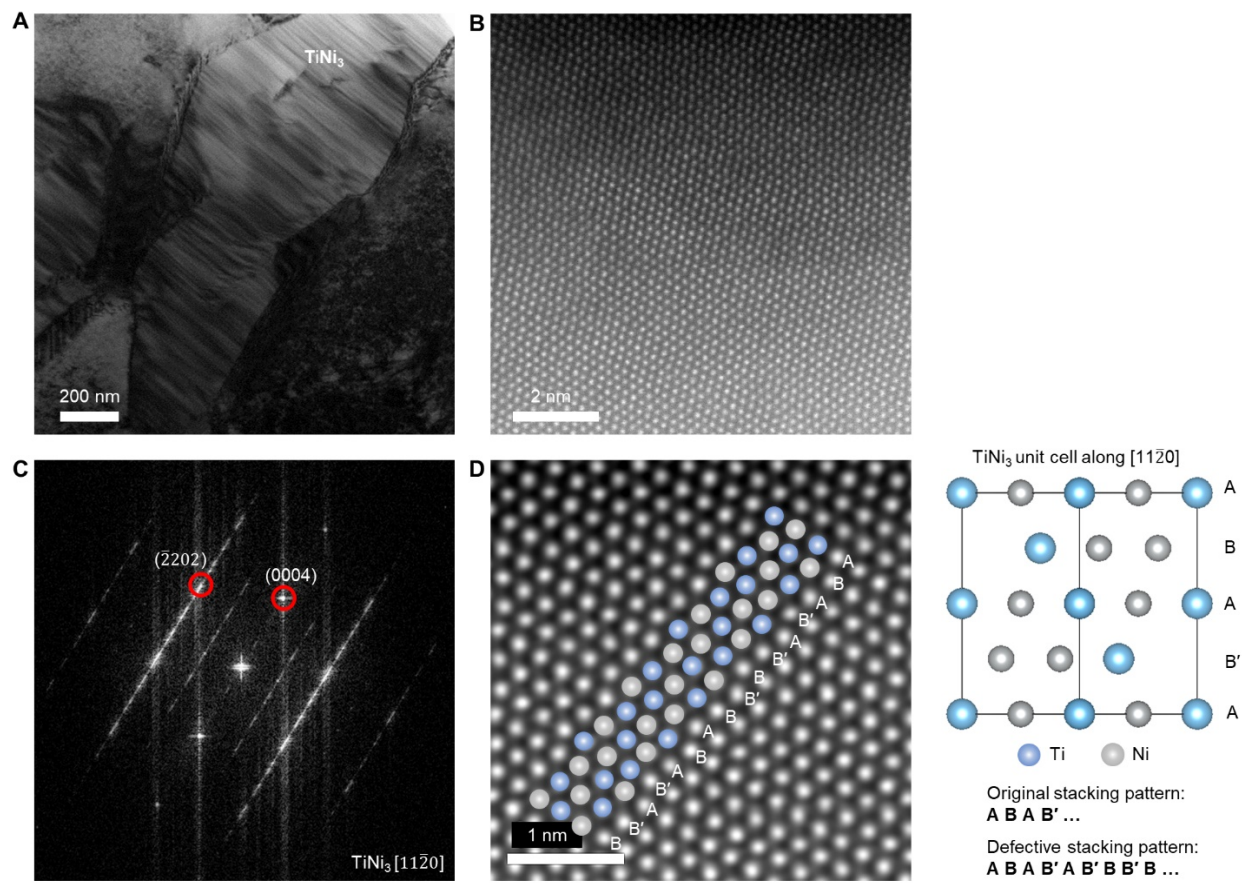

**Fig. S4. Stacking faults in the TiNi₃ phase.** (**A**) TEM diffraction-contrast image showing stacking faults within TiNi3 in the L-DED produced Ti–Ni nanocomposite alloys. (**B**) High-resolution HAADF image along TiNi3 [11$\bar{2}$0]. (**C**) FFT of (B). The streaks indicate the existence of the stacking faults. (**D**) Atomic-scale HAADF image with overlaid atomic models showing the defective stacking sequence. The regular stacking patterns of ABAB′… has been changed into ABAB′AB′BB′B… due to the stacking faults.

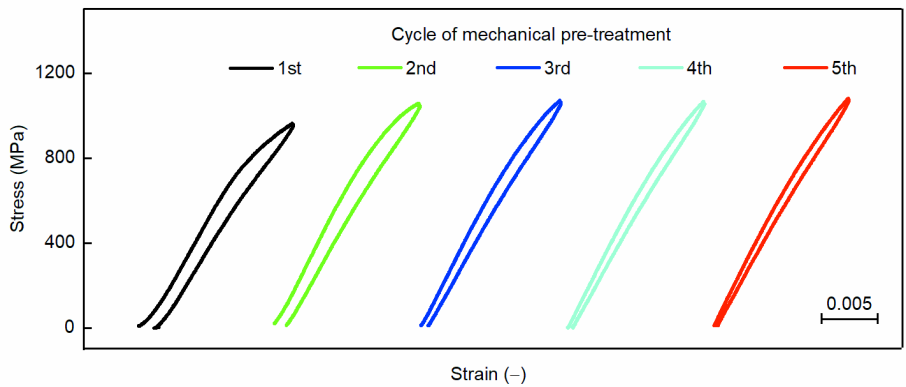

**Fig. S5. Mechanical pre-treatment for fully recoverable behaviors.** Mechanical pre-treatment is carried out on L-DED produced Ti–Ni nanocomposite alloys to attain reversibility. Initial compressive treatment during the first four cycles leads to a small residual strain upon unloading, a signature of progressing plastic deformation to relax local stress, which could come from $TiNi_3$ phase, and can also be facilitated by the defects such as stacking faults inside the $TiNi_3$ phase (seen in Fig. S4). Starting with the fifth cycle the increment in plastic deformation per cycle is negligible, indicating that the nanocomposite alloys has reached the shakedown state (*51, 52*); that is, beyond this point, further deformation of $TiNi_3$ phase is elastic due to the properly self-organized pre-straining and pre-stressing.

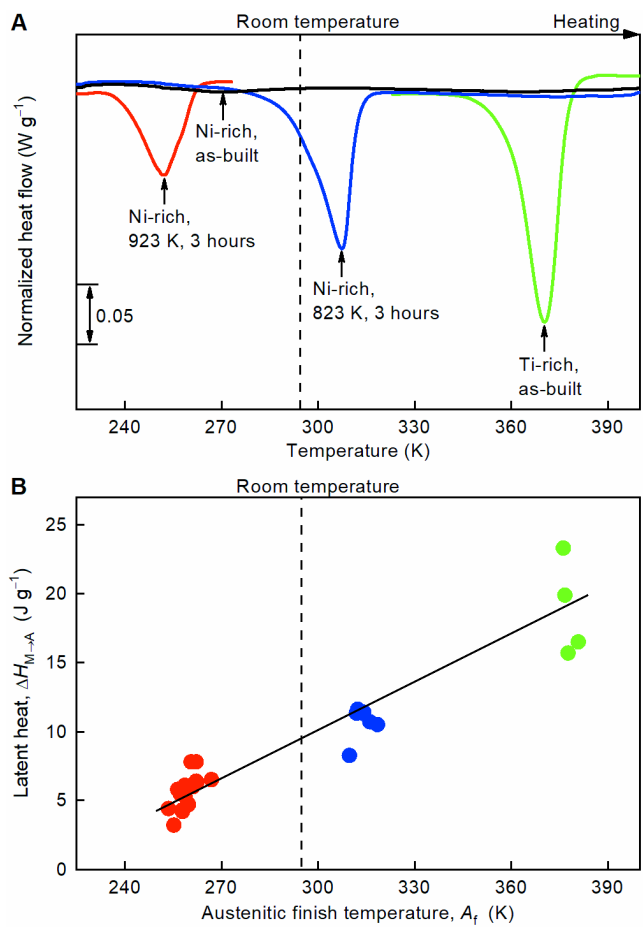

**Fig. S6. Tuned martensitic transformations toward room temperature.** (**A**) Differential scanning calorimetry thermo-grams of L-DED produced Ni-rich (51.5 at.% Ni) and Ti-rich (47.1 at.% Ni) Ti–Ni nanocomposite alloys after heat treatments (black: Ni-rich, as-built; red: Ni-rich, annealed at 923 K for 3 hours; blue: Ni-rich, annealed at 823 K for 3 hours; green: Ti-rich, as-built), displaying the phase transformation trend near or below room temperature. (**B**) Plot of austenitic finish temperature, $A_f$, versus endothermic latent heat, $\Delta H_{M \to A}$, displaying the wide range of the transformation temperatures and the latent heat. In (B), the solid line is a guide to the eye.

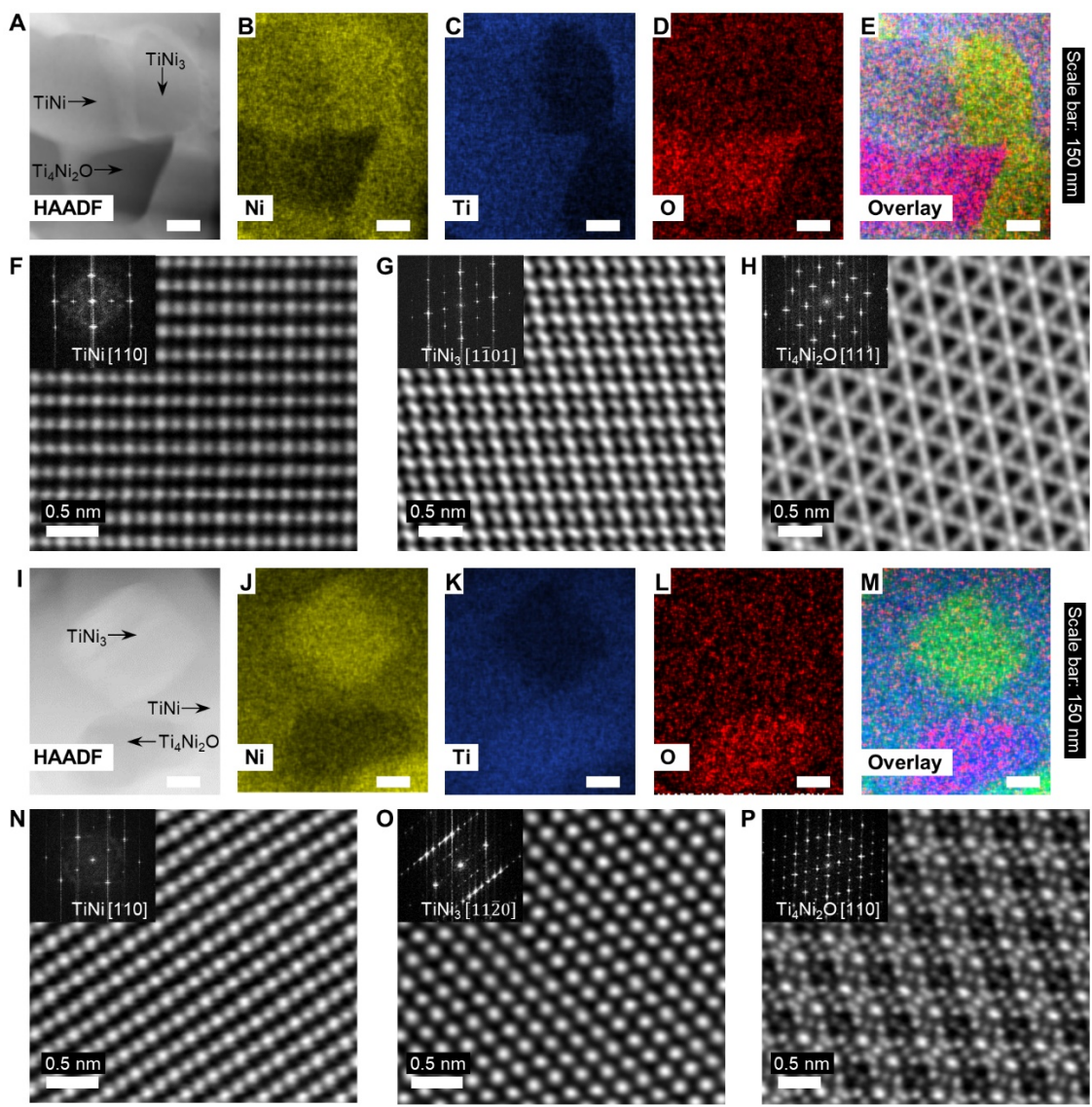

**Fig. S7. Energy-dispersive X-ray spectroscopy (EDS) analysis and HAADF imaging.** (A)–(P) L-DED produced Ti–Ni nanocomposite alloys after heat treatment of (A–H) 923 K for 3 hours and (I–P) 823 K for 3 hours. (**A**)–(**E**) and (**I**)–(**M**) HAADF-STEM image (A), (I), EDS mapping of Ni (B), (J), Ti (C), (K), and O (D), (L), and overlay EDS maps (E), (M). (**F**)–(**H**) and (**N**)–(**P**) Atomic-scale HAADF images of TiNi (F), (N), TiNi$_3$ (G), (O), and Ti$_4$Ni$_2$O phases (H), (P) with the corresponding fast Fourier transform (FFT) image and the zone axis at the top-left corner.

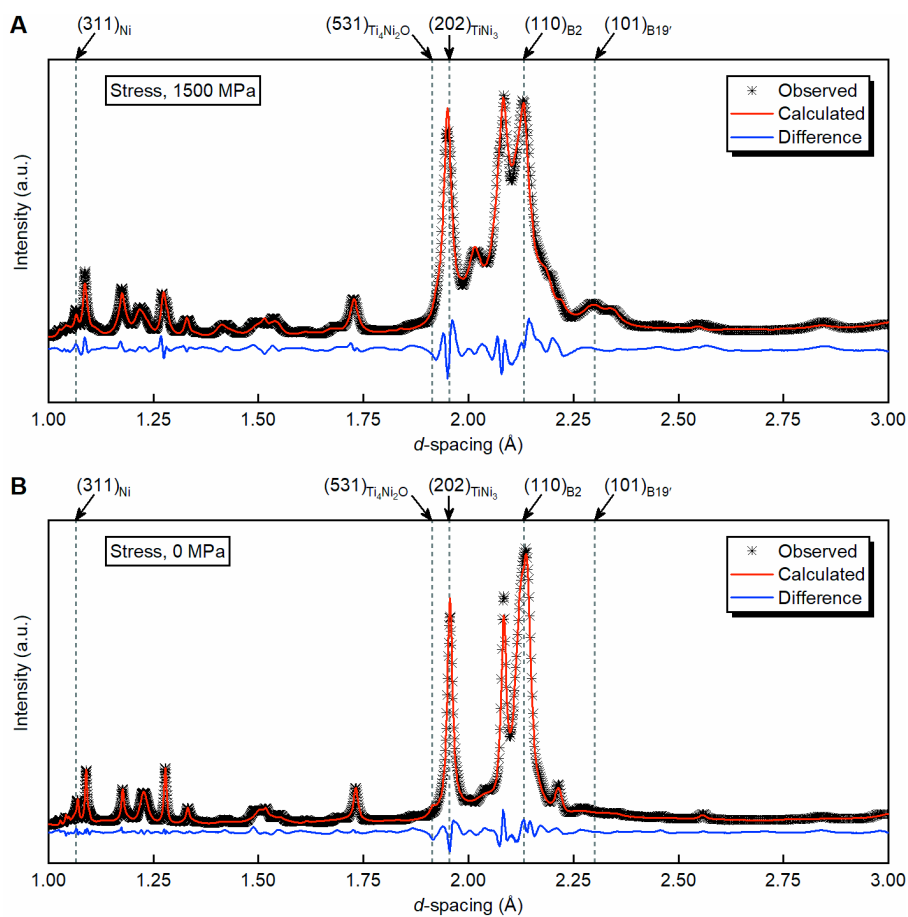

**Fig. S8. Quantitative Rietveld refinement.** (**A** and **B**) Rietveld refinement on high-resolution synchrotron X-ray diffraction patterns at a stress level of 1,500 MPa (A) and 0 MPa (B) to determine the present phases and their volume fractions. $TiNi_3$, $Ti_4Ni_2O$, and Ni have a volume fraction of 50.3±0.7%, 5.2±0.3%, and 0.8±0.1%, respectively. At a stress level of 0 MPa, TiNi-B2 has a volume fraction of 43.7±0.6%; at a stress level of 1,500 MPa, TiNi-B2 and TiNi-B19′ have a volume fraction of 10.9±0.6% and 33.7±0.6%, respectively. Roughly 50% of the nanocomposite being non-transforming precipitates is consistent with the measured latent heat of 5.6±1.3 J g$^{-1}$: they would correspond to 14.3±3.3 J g$^{-1}$ for the transforming fraction (TiNi) of the composite, well within the range of reported values (*5, 6, 50*).

**Table S1. Number of tested cycles for fatigue behavior of caloric materials reported in the literature.**

| Materials (at.%) | $T$ (K) | $\Delta E/E$ (%) | $\Delta\varepsilon/2$ * (%) | $N$ (–) | Reference | Caloric type & alloy category |
|---|---|---|---|---|---|---|
| $Ti_{48.5}Ni_{51.5}$ (P) | 294 | 6.9 | 1.8 | >1,000,000 | This work | |
| $Ti_{49.2}Ni_{50.8}$ (P) | 294 | 33.8 | 1.8 | >110,000 | This work | |
| $Ti_{49.2}Ni_{50.8}$ (S) | 313 | 18.3 | ≈1.5 | >10,000 | (53) | Elastocaloric, Ti–Ni-based |
| $Ti_{49.1}Ni_{50.9}$ (P) | 300 | 34.2 | 0.5 | >100,000 | (54) | |
| $Ti_{45.3}Ni_{50}V_{4.7}$ (P) | 298 | 27.5 | ≈1.9 | >5,000 | (55) | |
| $Ti_{54}Ni_{34}Cu_{12}$ (P) | 343 | 17.6 | ≈0.5 | >$10^7$ ** | (30) | |
| $Cu_{68}Zn_{16}Al_{16}$ (P) | 293 | 44.9 | 2.0 | >2,700 | This work | |
| $Cu_{59.1}Zn_{27}Al_{13.8}$ (S) | 343 | 15.6 | ≈2.5 | >10,000 | (53) | Elastocaloric, Cu-based |
| $Cu_{72}Al_{17}Mn_{11}$ (S) | 293 | 30.9 | 2.0 | >10,900 | (56, 57) | |
| $Ni_{54}Fe_{19}Ga_{27}$ (S) | 323 | 18.3 | ≈2.0 | >10,000 | (53) | |
| $Ni_{54}Fe_{19}Ga_{27}$ (P) | 298 | 42.8 | ≈1.5 | >100 | (58) | |
| $Ni_{50}Fe_{19}Ga_{27}Co_4$ (S) | 348 | 12.3 | ≈2.5 | >3,000 | (59) | |
| $Ni_{50.4}Mn_{27.3}Ga_{22.3}$ (P) | 327 | 44.0 | ≈1.4 | >100 | (60) | Elastocaloric, magnetic |
| $Ni_{48.4}Mn_{34.8}In_{16.8}$ (P) | 313 | 68.0 | ≈0.5 | >10 | (61, 62) | |
| $Ni_{51.3}Mn_{32.9}In_{15.5}B_{0.3}$ (P) | 303 | 25.6 | ≈1.1 | >150 | (62) | |
| $Ni_{45}Mn_{36.4}In_{13.6}Co_5$ (P) | 296 | 51.9 | ≈1.5 | >15 | (63) | |
| $Co_{50}Ni_{20}Ga_{30}$ (P) | 299 | 30.9 | ≈2.0 | >100 | (64) | |
| $Zn_{45}Au_{30}Cu_{25}$ | 298 | 8.0 | ≈2.2 | >100,000 | (31, 65) | Lattice-compatible |
| $LaFe_{11.6}Si_{1.4}$ (P) | 198 | 20.0 | – | 4 | (66) | |
| $LaFe_{11.6}Si_{1.4}$ (P; porous) | 198 | 11.1 | – | 800 | (66) | Magnetocaloric |
| $MnFe_{0.95}P_{0.595}B_{0.075}Si_{0.33}$ (P) | 278 | 2.8 | – | >10,000 ($\Delta T_{ad}$= 2.6 K) | (67, 68) | |
| $BaTiO_3$ (S) | 293 | 46.4 | – | 10 | (69) | Electrocaloric |
| $Ba(Zr_{0.2}Ti_{0.8})O_3$ (P) | 300 | 11.5 | – | 100,000 ($\Delta T_{ad}$= 0.2 K) | (70) | |
| $Fe_{49}Rh_{51}$ (P) | – | – | – | 50–100 | (71) | Barocaloric*** |

$T$: Test temperature, $\Delta E/E$: dissipated fraction of input energy, $\Delta\varepsilon/2$: strain amplitude, and $N$: sustained cycles. "S" in parenthesis stands for single crystal and "P" stands for polycrystal. "Lattice-compatible" refers to the alloy whose lattice parameters meet the mathematical relation that gives rise to a high degree of compatibility between phases (31). The $\Delta E/E$ can be evaluated for elastocaloric materials from stress–strain ($\sigma$–$\varepsilon$) plot, for magnetocaloric materials from magnetization–applied magnetic field ($M$–$H$) plot, and for electrocaloric materials from polarization–applied electric field ($P$–$E$) plot.

\* An exact value is listed when the value is explicitly stated in the reference. Otherwise, an approximate value is extracted from the reference.

\*\* The materials are in thin film form with a thickness of 18 µm, and there are no cooling data reported in that reference.

\*\*\* In the reference, the cycle number is by indirect methods.